\documentclass[twocolumn]{aastex61}
\usepackage[T1]{fontenc}
\usepackage{apjfonts} 
\usepackage{systeme}
\usepackage{listings}
\usepackage{booktabs} 
\usepackage{multirow}
\usepackage{amsmath, graphicx}

\shorttitle{ML methods for automated ISO classification with LSST}
\shortauthors{Cloete et al.}

\begin{document}

\def\thefootnote{*}\footnotetext{These authors contributed equally to this work}\def\thefootnote{\arabic{footnote}}

\title{Machine Learning Methods for Automated Interstellar Object Classification with LSST}

\author[0000-0003-2586-2697]{Richard~Cloete 1$^*$}
\affiliation{Harvard-Smithsonian Center for Astrophysics, 60 Garden St., MS 15, Cambridge, MA 02138, USA}

\author[0000-0002-5396-946X]{Peter~Vere\v{s} 1$^*$} 
\affiliation{Harvard-Smithsonian Center for Astrophysics, 60 Garden St., MS 15, Cambridge, MA 02138, USA}

\correspondingauthor{Richard Cloete}
\email{richard.cloete@cfa.harvard.edu}

\author{Abraham Loeb}
\affiliation{Harvard-Smithsonian Center for Astrophysics, 60 Garden St., MS 15, Cambridge, MA 02138, USA}

\keywords{Interstellar Objects, LSST, Vera Rubin Observatory, Machine learning, Short-arc Orbit Determination, Near-Earth Objects, Astronomy, Solar System}

\begin{abstract}
{The Legacy Survey of Space and Time (LSST), to be conducted with the \textit{Vera C. Rubin} Observatory, is poised to revolutionize our understanding of the Solar System by providing an unprecedented wealth of data on various objects, including the elusive interstellar objects (ISOs). Detecting and classifying ISOs is crucial for studying the composition and diversity of materials from other planetary systems. However, the rarity and brief observation windows of ISOs, coupled with the vast quantities of data to be generated by LSST, create significant challenges for their identification and classification.}
{This study aims to address these challenges by exploring the application of machine learning algorithms to the automated classification of ISO tracklets in simulated LSST data.}
{We employed various machine learning algorithms, including random forests (RFs), stochastic gradient descent (SGD), gradient boosting machines (GBMs), and neural networks (NNs), to classify ISO tracklets in simulated LSST data.}
{Our results demonstrate that GBM and RF algorithms outperform SGD and NN algorithms in accurately distinguishing ISOs from other Solar System objects. RF analysis shows that many derived Digest2 values are more important than direct observables (right ascension, declination, and magnitude) in classifying ISOs from the LSST tracklets. The GBM model achieves the highest precision, recall, and F1 score, with values of 0.9987, 0.9986, and 0.9987, respectively.}
{These findings lay the foundation for the development of an efficient and robust automated system for ISO discovery using LSST data, paving the way for a deeper understanding of the materials and processes that shape planetary systems beyond our own. The integration of our proposed machine learning approach into the LSST data processing pipeline will optimize the survey's potential for identifying these rare and valuable objects, enabling timely follow-up observations and further characterization.}
\end{abstract}

\newcommand{\hl}[1]{#1}

\section{Introduction}
\label{sec:intro}

The \textit{Vera C. Rubin} Observatory, formally called the Large Synoptic Survey Telescope, is set to revolutionize the field of astronomy when it begins operations in 2025~\citep{ivezic2019lsst} with its unprecedented wide-field survey capabilities. The survey will generate a vast amount of data, enabling research in diverse areas of astrophysics, including the discovery and characterization of new Solar System objects~\citep{Schwamb23} such as interstellar objects \citep[ISOs;][]{jones2009solar}. To date, only two ISOs have been discovered (excluding interstellar meteors): 1I/`Oumuamua and 2I/Borisov.

1I/`Oumuamua was discovered coincidentally as an unknown object moving at a high apparent rate of motion. Due to its high near-Earth object (NEO) Digest2 score \citep{keys2019digest2}, the object was posted to the Near-Earth Object Confirmation Page\footnote{\url{https://minorplanetcenter.net//iau/NEO/toconfirm_tabular.html}} (NEOCP) of the Minor Planet Center\footnote{\url{https://minorplanetcenter.net/}} (MPC), which enabled rapid follow-up observations from multiple sites around the world. Within a few days, the heliocentric orbit was deemed to be undoubtedly hyperbolic. At the same time, 1I/`Oumuamua was classified as an NEO due to its perihelion distance of less than 1.3 astronomical units (AU).

After its discovery, astronomers noticed 1I/`Oumuamua's peculiar physical properties, such as its extremely elongated shape \citep{meech2017brief,Jewitt17,Bannister17,Masiero17,Knight17,Ye17,Fraser18,Bolin18,fitzsimmons2018spectroscopy,Ma19} and its lack of cometary activity \citep{Trilling18}. These characteristics sparked an intense debate about its origin and composition \citep{Flekk19,Hoang20,Curran21,Jackson21,Si22,Flekkoy22,Loeb23}. The motion of 1I/`Oumuamua has not yet been fully explained, with several studies proposing different explanations for its nongravitational acceleration \citep{Micheli18,Bialy18,Rafikov18,Loeb22,Bergner23,Hoang23}.

Conversely, 2I/Borisov was discovered as a new comet~(see \citealt{Je23} and references therein), and like other new comet discoveries, the object was posted to the MPC's Possible Comet Confirmation Page\footnote{\url{https://data.minorplanetcenter.net/iau/NEO/pccp_tabular.html}} (PCCP).\ This again enabled rapid follow-up observations and early orbit determination, which proved that the comet has a highly hyperbolic orbit with respect to the Sun. Therefore, the second ISO was also a coincidental discovery, with its cometary activity drawing the attention of astronomers.

However, if an ISO's orbit is not known and it is observed as a short intra-night object, it can be easily missed or misclassified as another type of object. This emphasizes the need for an automated approach for efficiently detecting and classifying ISOs. Additionally, the vast quantities of data that will be produced by LSST, especially in the first few years as new ``background" objects --- particularly main belt asteroids (MBAs) --- are discovered, will further impede our ability to identify possible ISOs.

This will be particularly challenging in the first few years of LSST operations as most of the objects seen will be new discoveries, dominated by small MBAs.
Traditional methods of data analysis cannot feasibly handle such massive datasets in a timely manner. This is where machine learning (ML) becomes crucial. Machine learning techniques are powerful tools for automatically processing and analyzing large volumes of data, identifying patterns, and making predictions with high accuracy and efficiency.

In the context of ISO detection and classification, ML algorithms can be trained to recognize the unique motion characteristics of ISOs amidst a vast sea of Solar System objects. These algorithms can rapidly sift through the data, flagging potential ISOs for further analysis and follow-up observations. By automating the identification process, ML not only accelerates the discovery of new ISOs but also improves the reliability and consistency of detections.

Therefore, to address the challenge of automatically flagging candidate ISOs for follow-up observation, we explored and evaluated several state-of-the-art ML algorithms for automated tracklet classification, including random forests~\citep[RFs;][]{breiman2001random}, stochastic gradient descent \citep[SGD;][]{bottou2010large}, gradient boosting machines \citep[GBMs;][]{friedman2001greedy}, and neural networks~\citep[NNs;][]{lecun2015deep}. By comparing the performance of these algorithms on simulated LSST data, we aim to identify the most effective approach to accurately classifying tracklets and distinguishing ISOs from other Solar System objects. The results of this study will facilitate the development of a robust and efficient automated system for ISO discovery and characterization using LSST data.

 \section{The search for interstellar objects}
\label{s:related-work}

The existence of ISOs has been theorized for decades. In the early stages of the Solar System, a large quantity of planetesimals and debris was ejected due to the instability of the dynamical system and the gravity of the giant planets~\citep{Charnoz03,Bottke05,Raymond18,Raymond20}. Like small planetesimals, larger bodies such as planets (so-called free-floating planets) can be ejected from their parent systems~\citep{Scholz12,Pena12,Miret22}.

Several studies derived upper estimates of the spatial density of ISOs around the Sun before the discovery of 1I/`Oumuamua and 2I/Borisov~\citep{Torbett86,McGlynn89,Sen93,Jewitt03,Moro09,Cook16}. Some estimates relied on ongoing Solar System surveys, their pointing data and depth, such as LINEAR~\citep{Francis05} or Pan-STARRS~\citep{engelhardt2017observational}. However, without a single ISO discovered, the estimates varied by orders of magnitude, between $10^{-9}\,au^{-3}$ and $2.4\times10^{-2}\,au^{-3}$ for a 1-km ISO.

The discovery of 1I/`Oumuamua and 2I/Borisov allowed a better constraint to be placed on the spatial density of both active and inactive ISOs:~\cite{Jewitt17} predicted $1\times au^{-3}$ `Oumuamua-like objects closer than Neptune at any given time, while~\cite{Levine21} predicted the 3-sigma number density of similar sized ISOs between $2 \times 10^{-4}au^{-3}$ and $0.8au^{-3}$.~\cite{Do18} derived the number density to be similar to $0.2au^{-3}$ and~\cite{Zwart18} derived a number density of $0.0040-0.24au^{-3}$. Meanwhile,~\cite{bolin2020characterization} used the data from both discovered ISOs and derived a density of $\sim1au^{-3}$ \hl{for slightly larger, 250-meter ISOs, and determined the slope of the size-frequency distribution of ISOs $N(>D)$, in terms of actual size (km), to be $-3.38\pm1.18$}.

Other works explored the potential of the LSST for discovering and characterizing ISOs and other unique Solar System objects. LSST's unprecedented limiting magnitude of about $+24.5$ in g-band\footnote{\url{https://www.lsst.org/scientists/keynumbers}} and sky-coverage of approximately half of the sky during the survey, offer a unique opportunity to detect the interstellar interlopers.

\cite{Hoover22} predicted that LSST will detect 1-3 ISOs of 1I/`Oumuamua's size and properties per year.
\cite{marvceta2023synthetic} estimated that LSST could detect between 0 and 70 ISOs per year, depending on their albedo and size-frequency distribution, thus covering a wide range of possibilities. 

These studies collectively demonstrate the growing interest in the detection and characterization of ISOs, as well as the potential of the LSST to significantly advance this field. However, they fail to directly address the challenge of identifying ISOs in LSST data with high confidence \hl{and in a timely manner, which is critical for effective characterization. While ISO candidates can eventually be identified post-processing or through analysis by the MPC, rapid identification is crucial for follow-up observations, especially for objects on transient trajectories like 1I/`Oumuamua, where early detection would have allowed for a more detailed characterization.}
By leveraging synthetic LSST data provided by the \textit{Rubin} Science Platform (RSP)\footnote{\url{https://data.lsst.cloud}}, we aim to contribute to this ongoing research effort by employing ML algorithms as a means to automatically flag potential ISO candidates for follow-up observation among the large quantities of daily alerts. The next sections describe our methodology, model development, and evaluation.
 \section{Methodology}
\label{s:methodology}

LSST Solar System products will be delivered in several forms: real-time alerts will provide large amounts of individual transients (400,000 to 5 million per night\footnote{\url{https://dmtn-102.lsst.io/DMTN-102.pdf}}), and a daily batch of tracklets will be shared with the \hl{MPC}. Tracklets are short sequences of observations of the same object observed two or more times per night, and they can be used to determine the object's approximate orbit (or range of potential orbits) and to classify them as NEOs, MBAs, or other types of objects~\citep{denneau2013pan,keys2019digest2}, such as ISOs.

\hl{LSST will submit to the MPC both individual tracklets and linkages of tracklets that represent an object that has been observed on three or more nights \citep{Holman18, Heinze22}, but they will not submit unlinked individual detections.}
Additionally, LSST will identify known objects and submit them with their designations to the MPC. The LSST catalog of derived orbits will also include photometric and light-curve information, although at present we have no plans to make use of such data. Although intra-night linking has been widely used by surveys such as WISE Moving Object Pipeline Subsystem \citep{Dailey10}, Pan-STARRS Moving Object Processing System \citep{Denneau13}, or Zwicky Transient Facility’s Moving Object
Discovery Engine \citep{Masci19}, the efficiency of the inter-night linking by LSST, which currently utilizes the HelioLinC method \citep{Holman18, Heinze22}, has not yet been proven for hyperbolic orbits. Consequently, the ability to link hyperbolic orbits and thus identify ISOs remains uncertain.

Our work focuses on analyzing the intra-night tracklets of unidentified objects as observed by LSST. Specifically, we used directly reported LSST ``observables'' --- the right ascension, declination, and magnitude at a given epoch --- and derived values such as the sky-plane velocity and motion direction and the apparent position with respect to the opposition. We also employed Digest2, using its output as additional input for our ML models. In this section we describe our data sources, processing, feature selection process, and their derived values employed by ML models for the purpose of automatically classifying ISO tracklets. 

\subsection{LSST input data}
\label{s:lsst_input_data}

We utilized the LSST DP0.3 Solar System object simulation data, which contains a 1-year ($dp03\_catalogs\_1yr$) and 10-year ($dp03\_catalogs\_10yr$) quasi-realistic distribution of LSST pointings with simulated detections of Solar System objects, including both real (discovered) objects from the Minor Planet Center Orbit Database (MPCORB) and the synthetic Solar System model \citep{Grav11}. The simulations collectively contain more than 13 million synthetic orbits, including approximately 12,000 hyperbolic (ISO) orbits. Data are distributed as Astronomical Data Query Language (ADQL) databases available through the RSP\footnote{\url{https://data.lsst.cloud/}}, each with four tables: DiaSources (simulated astrometric and photometric measurements for detected Solar System objects), MPCORB (catalog of real and synthetic orbits), SSObject (table of LSST-linked Solar System objects), and SSSource (Solar System source information corresponding to specific difference image detections).

\begin{table}[ht!]
\centering
\caption{Number of tracklets  and objects in the 1-year LSST simulation. The table shows the synthetic population of more than 11 million input orbits taken from \citet{Grav11}}
\label{tab:tracklets_1y}
\begin{tabular}{l|r|r}
\hline
\textbf{Object Type} & \textbf{Tracklets} &  \textbf{Objects} \\ 
\hline
    Near-Earth Objects (NEO) & 21,723 & 3,809\\ 
    Main-Belt Asteroids (MBA) & 5,749,986 & 844,656\\ 
    Centaurs & 9,795 & 998 \\ 
    Interstellar Objects (ISO) & 3,306 & 800 \\ 
    Comets & 5,608 & 740 \\ 
    Scattered Disk (Distant TNO) & 17,557 & 2,224 \\ 
    Trans-Neptunian Objects (TNO) & 88,410 &11,022 \\ 
    Trojans & 179,321  & 33,412\\ 
\hline
\end{tabular}

\end{table}

\begin{figure}[ht]
   \includegraphics[width=\linewidth]{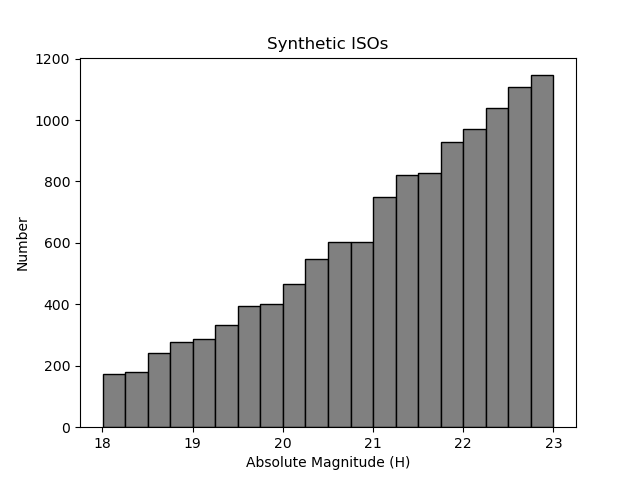}
    \caption{Histogram of absolute magnitudes of synthetic ISOs used in this work.}
    \label{fig:ISO_H}
\end{figure}

Table~\ref{tab:tracklets_1y} illustrates the number of detections by category in the 1-year LSST simulation data.
Although the synthetic model should realistically balance the ratio of different orbital types based on object size, the ISO population is significantly exaggerated in quantity\hl{~\citep{Grav11}}. For this reason, we downloaded all data from the 1-year dataset and only the ISO detections from the 10-year dataset, thus boosting the total number of ISO samples from 3,306 to 14,151. Figure ~\ref{fig:ISO_H} shows the size-frequency distribution of synthetic ISO orbits: synthetic ISOs have H in a range of 18-23, representing roughly objects of a size of between 100 meters to 1 kilometer \hl{for an assumed albedo of 0.1}.

\subsection{Boosting the sample count of synthetic ISO tracklets}
\label{ss:pseudo_isos}

Our initial dataset had about 6 million tracklets, of which only 14,151 (0.24\%) were ISOs.\ The remaining tracklets were other celestial objects --- a highly imbalanced dataset.

Highly imbalanced data in classification problems significantly impacts the performance and reliability of predictive models~\citep{sun2009classification,japkowicz2002class,chawla2002smote,he2009learning}. In scenarios where the distribution of classes is skewed, traditional algorithms can exhibit a bias toward the majority class, leading to an inadequate representation and misclassification of the minority class. This imbalance can distort the learning process, as models may prioritize minimizing errors on the more prevalent class, consequently neglecting the rare yet potentially more critical class. 

To address this imbalance, we sought to boost the number of ISO samples and then randomly sample an equal number of non-ISOs to form a new, balanced dataset.
We downloaded the heliocentric ISO orbits from the $dp03\_catalogs\_10yr$ simulation, which contain 12,148 Keplerian hyperbolic orbits. 

To significantly increase the number of ISO tracklets, we created a simplistic LSST-like pseudo-survey. First, we propagated 12,148 ISO orbits to the initial epoch of the 10-year survey (to the local midnight) and then to a second epoch one our later, thus creating two-detection 1-hour tracklets for each orbit. This approach allowed us to generate a larger number of synthetic ISO tracklets that closely resembled the expected observations from the LSST survey.

\begin{figure}[ht]
    \includegraphics[width=\linewidth]{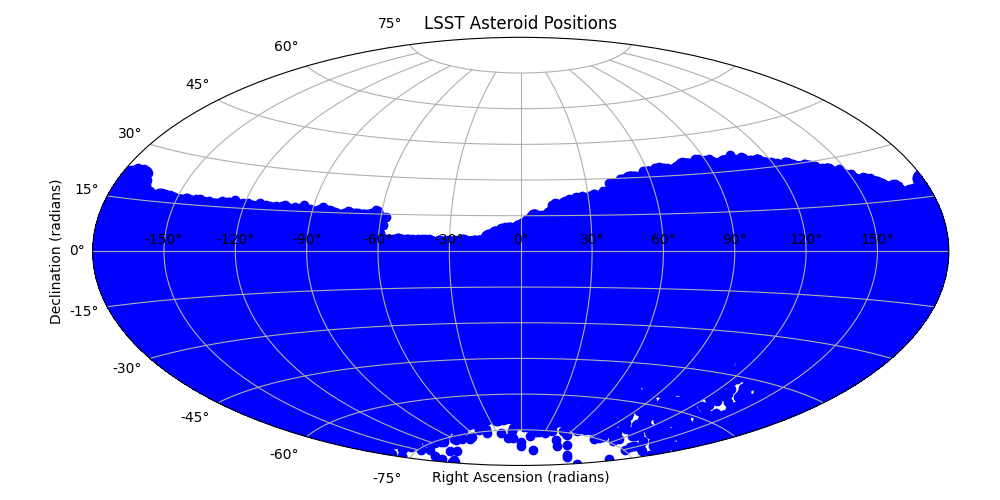}
    \caption{Hammer-Aitoff sky-plane projection of synthetic positions of Solar System objects in downloaded LSST data. }
    \label{fig:lsst_hammer}
\end{figure}

Subsequently, we computed the ephemerides for each day over the next 10 years, using the following constraints to simulate a detection: 

\begin{itemize}
    \item a limiting V-band magnitude of 24.5,
    \item a minimum apparent lunar elongation of 90 degrees,    \item a minimum solar elongation of 60 degrees,
    \item a minimum object altitude of 20 degrees,
    \item allowing only detections with a negative declination or with a declination greater than 0 but an ecliptical latitude of less than 10 degrees, mimicking the LSST survey area as seen in Fig.~\ref{fig:lsst_hammer}.
\end{itemize}

We considered a tracklet valid when the same object fulfilled the mentioned constraints and was detected twice on the same night. To create a dataset independent from the original one but maintaining the same orbits, we shuffled the absolute magnitude $H$ values, ensuring that the size-frequency distribution remained the same. This custom synthetic dataset of ISO tracklets was then added to the LSST-generated tracklets. The total number of generated ISO tracklets is displayed in Table~\ref{t:low_fidelity_ISO}.

\begin{table}
\begin{center}
    \caption{Low-fidelity simulation of ISO tracklets in 10 years of the LSST survey.}
\begin{tabular}{l|c|r}
        \hline
        \textbf{Simulation} &  \textbf{Tracklets} &  \textbf{Orbits}\\
        \hline
        Original ISO & 223949 &1218\\
        Shuffled H & 194642 &  864\\
        \hline
        Total number  &  210096 & 1739\\
        \hline
        \end{tabular}
    \label{t:low_fidelity_ISO}
    \end{center}
\end{table}

\subsection{Digest2}
\label{ss:digest2}

Digest2~\footnote{\url{https://bitbucket.org/mpcdev/digest2}} is a robust and efficient short-arc orbit classifier for small Solar System bodies, primarily utilized for the identification and prioritization of NEO candidates for follow-up observations~\citep{keys2019digest2}.  
Digest2 operates by analyzing tracklets, employing statistical methods for motion analysis and initial orbit determination. Each object processed by Digest2 is assigned a ``D2'' score (often referred to as the NEO score) ranging from 0 to 100, representing a pseudo-probability that a tracklet belongs to an NEO. In addition to the D2 NEO score, Digest2 outputs scores for 14 additional orbit classes (see Table 8 in~\citealt{keys2019digest2}). There are two independent scores for each class, ``raw'' is the score with respect to the entire model population as if all objects have been discovered, and ``noid'' score is the score with respect to the undiscovered portion of a given population \citep{keys2019digest2}.

Though Digest2 is regularly used to tag NEO candidates, the code has not been fully utilized to identify objects of other orbital categories despite its ability to do so. Therefore, we sought to build on Digest2 by exploring whether its output values could aid in the task of classifying ISOs.

One of the key questions we had was whether the output from Digest2 would be important in classifying ISOs. That is, would the orbital categories output by Digest2 serve as important input features for our ML models when determining whether a tracklet belongs to an ISO?

\begin{figure}[ht!]
    \includegraphics[width=\linewidth]{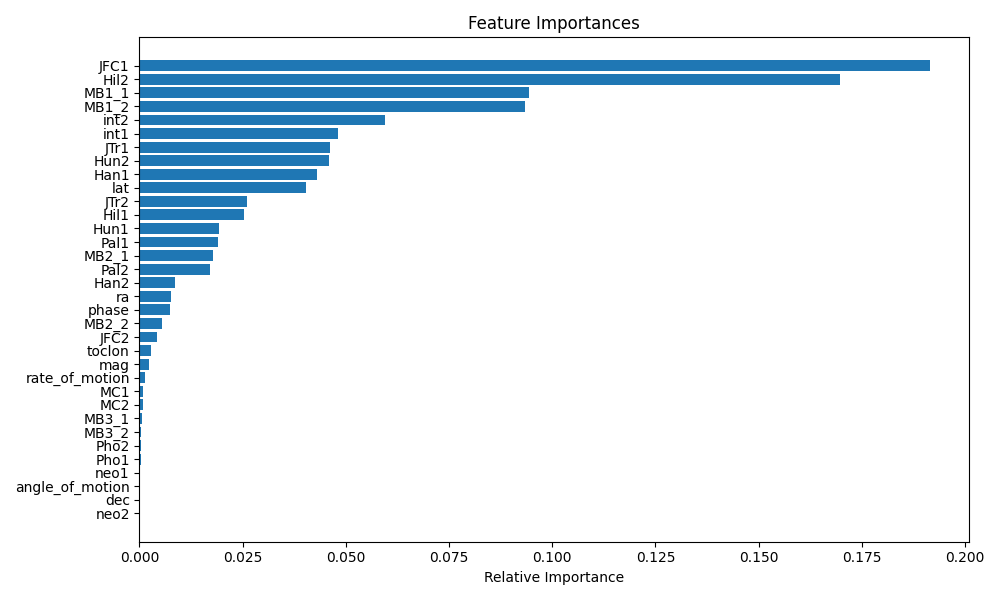}
    \caption{Digest2 output values ranked in the top nine our of ten features identified as important for ISO classification.}
    \label{fig:rf-features}
\end{figure}

To explore this, we employed the RF algorithm to generate feature importances, which measure the contribution of each feature to the model's predictive performance~\citep{louppe2013understanding}. During our analysis, we noticed that indeed many of the features identified as having high importance were produced by Digest2, rather than those that were derived or obtained directly from the simulated LSST data. This is evident from Fig.~\ref{fig:rf-features}, which shows that nine out of the ten most important features identified using the RF method were from Digest2.
Among the highest-ranked features were those related to Jupiter-family comets (``raw'' Digest2), Hildas (``noid'' Digest2), inner main belt, and ``Interesting'' categories. Importantly, these features often relate to celestial objects with high eccentricity (Jupiter-family comets and Interesting) or to distant objects that have slow motion (Jupiter-family comets and Hildas). This suggests that these characteristics likely play a key role in identifying ISOs, which are known to have highly eccentric (hyperbolic) orbits and can have inclinations at any angle. Further explanation of the Digest2 algorithm and its output can be found in \cite{keys2019digest2}.

Having confirmed that Digest2 output will serve as important input for our models, the next steps involved preprocessing our data and integrating the Digest2 output, which we describe next.

\subsection{Data preprocessing}
\label{ss:preprocessing}

\begin{table*}[!t]
\caption{Data sources and corresponding columns used in the analysis. The LSST columns represent the raw data provided by the simulated LSST observations, while the derived columns are computed from the LSST data. The Digest2 columns are the output scores from the Digest2 short-arc orbit classifier for various orbital categories defined in \citep{keys2019digest2} with the suffix ``1'' representing the ``raw'' and ``2'' the ``noid'' score.}
\centering
\begin{tabular}{l|l}
\hline
\textbf{Data Source} & \textbf{Columns} \\ \hline
LSST & $\alpha$, $\delta$, mag, lat, phase, orbtype \\ 
Derived & toclon, rate\_of\_motion, angle\_of\_motion \\ 
\multirow{3}{*}{Digest2} & int1, int2, neo1, neo2, MC1, MC2, Hun1, Hun2, Pho1, Pho2, MB1\_1, MB1\_2, Pal1, Pal2  \\ 
                         & Han1, Han2, MB2\_1, MB2\_2, MB3\_1, MB3\_2, Hil1, Hil2, JTr1, JTr2, JFC1, JFC2 \\  
                           \hline
\end{tabular}
\label{tab:data_sources}

\end{table*}

As mentioned in Sect. \ref{s:lsst_input_data}, we downloaded all of the one-year LSST DP0.3 Solar System object simulation data and only ISOs from the ten-year simulation.

We then processed the downloaded individual detections using known object designations. Each detection was given a custom tracklet ID (\texttt{trksub}), which we derived by combining the modified mpcDesignation with the integer part of the \hl{modified Julian date (MJD)} plus a constant offset of 0.5. Each $trksub$ has a length of 12 characters. 

The data were then filtered to ensure that each object had a minimum of two detections per night, and then grouped by their \texttt{trksub} to form tracklets. That is, we define a tracklet as the same object observed two or more times per night.
\hl{The tracklets were labeled by the nature of their orbits: ISOs and everything else. In our work, we assumed the tracklet linking efficiency is ideal, and we did not account for false tracklets (mislinked objects or object-noise linkages) such as discussed in \citet{Veres17b}.}

The apparent magnitude in the observed filter band was converted to an approximate V-band magnitude (vmag) using a conversion factor specific to each filter\footnote{\url{https://www.minorplanetcenter.net/iau/info/BandConversion.txt}}. The detections were then grouped by their unique \texttt{trksub} identifier.

For each tracklet (detection group), we selected the first and last detections (skipping those where the MJDs were the same) and derived quantities such as the rate of motion, position angle, ecliptical latitude, opposition-centered ecliptical longitude, and solar elongation. The \texttt{orbtype} column was created to indicate the class of the celestial object, with a value of 0 or a 1 assigned to each tracklet based on the leading characters of the modified mpcDesignation column, which we modified to start with an ``I'' for ISOs or a ``1'' for all other obit classes. 
The processed LSST data were compiled into a new dataset and merged with the pseudo ISO tracklets described in Sect. \ref{ss:pseudo_isos}.

\hl{The next step involved using Digest2. Digest2 ingests tracklets, represented by observational data, in so-called MPC1992 format\footnote{\url{https://www.minorplanetcenter.net/iau/info/OpticalObs.html}}}. 
We converted previously prepared LSST observations to MPC1992 format, particularly the 12-character \texttt{trksub}, \texttt{epoch}, \texttt{right} \texttt{ascension} ($\alpha$), \texttt{declination} ($\delta$), \texttt{magnitude}, and \texttt{band} into the MPC1992 format, with the LSST observatory code $X05$\footnote{\url{https://minorplanetcenter.net/iau/lists/ObsCodesF.html}}. The resulting format was generated with full-precision in $\alpha$, $\delta$, magnitude and the epoch. For this, the Digest2 source code was slightly altered so that the program could ingest 12-character \texttt{trksubs}. We computed 13 Digest2 parameters in both ``raw'' and ``noid'' modes for each of our synthetic tracklets, resulting in 26 features, and concatenated the output with our dataset (see Sect. \ref{s:lsst_input_data} for the derived quantities).

Next, we removed duplicate entries based on the \texttt{trksub} column. During this data cleaning phase, we noticed that some detections had apparent magnitudes fainter than the limiting magnitudes\footnote{\url{https://www.lsst.org/scientists/keynumbers}} and we therefore removed a few percent of the detections. The remaining data were sorted in ascending order by \texttt{trksub} and MJD to establish a consistent sequence of observations. Irrelevant columns, including \texttt{trksub} and MJD, were dropped, and the other columns were renamed for clarity.
Our final dataset is described in Table~\ref{tab:data_sources}.

 \section{Model training and evaluation}
\label{s:evaluation}

Having identified the key features and completed data preprocessing, we split the dataset into training, testing, and validation subsets using a two-step process. The data were first divided into training (80\%) and temporary (20\%) sets using a fixed random seed of 42 for reproducibility. The temporary set was then further split equally into testing and validation sets using the same random seed.

Within each subset, we separated the target variable (\texttt{orbtype}) from the feature variables. The target variable represented the class or category we aimed to predict (i.e., non-ISO vs. ISO), while the feature variables encompassed all the remaining columns that would be used as input to the ML models.

\subsection{Model selection}
\hl{
Using these data, we trained and evaluated several ML models for ISO detection: GBMs~\citep{friedman2001greedy}, RFs~\citep{breiman2001random}, SGD~\citep{bottou2010large}, and NNs~\citep{lecun2015deep}.

The GBM algorithm is an ensemble learning method that builds a series of weak learners, typically decision trees, sequentially to correct errors made by previous models. GBM is known for its high predictive accuracy and ability to handle complex, nonlinear relationships in data.

The RF algorithm is another ensemble method that constructs multiple decision trees and combines their outputs for prediction. RF is particularly effective at reducing overfitting through its use of bagging and random feature selection, making it robust across various types of datasets.

The SGD algorithm optimizes the model by updating weights incrementally using randomly chosen data points, making it computationally efficient for large datasets. Although it requires careful tuning, SGD can quickly converge to good solutions in high-dimensional spaces where other models may struggle.

Neural networks, with their layered architecture, are capable of capturing complex, nonlinear relationships. NNs learn hierarchical feature representations through back-propagation, making them particularly suited for tasks with intricate patterns, though they often require more data and computational resources to reach optimal performance.
}

\subsection{Evaluation results}
 
\begin{figure*}[!ht]
    \centering
    \includegraphics[width=0.45\linewidth]{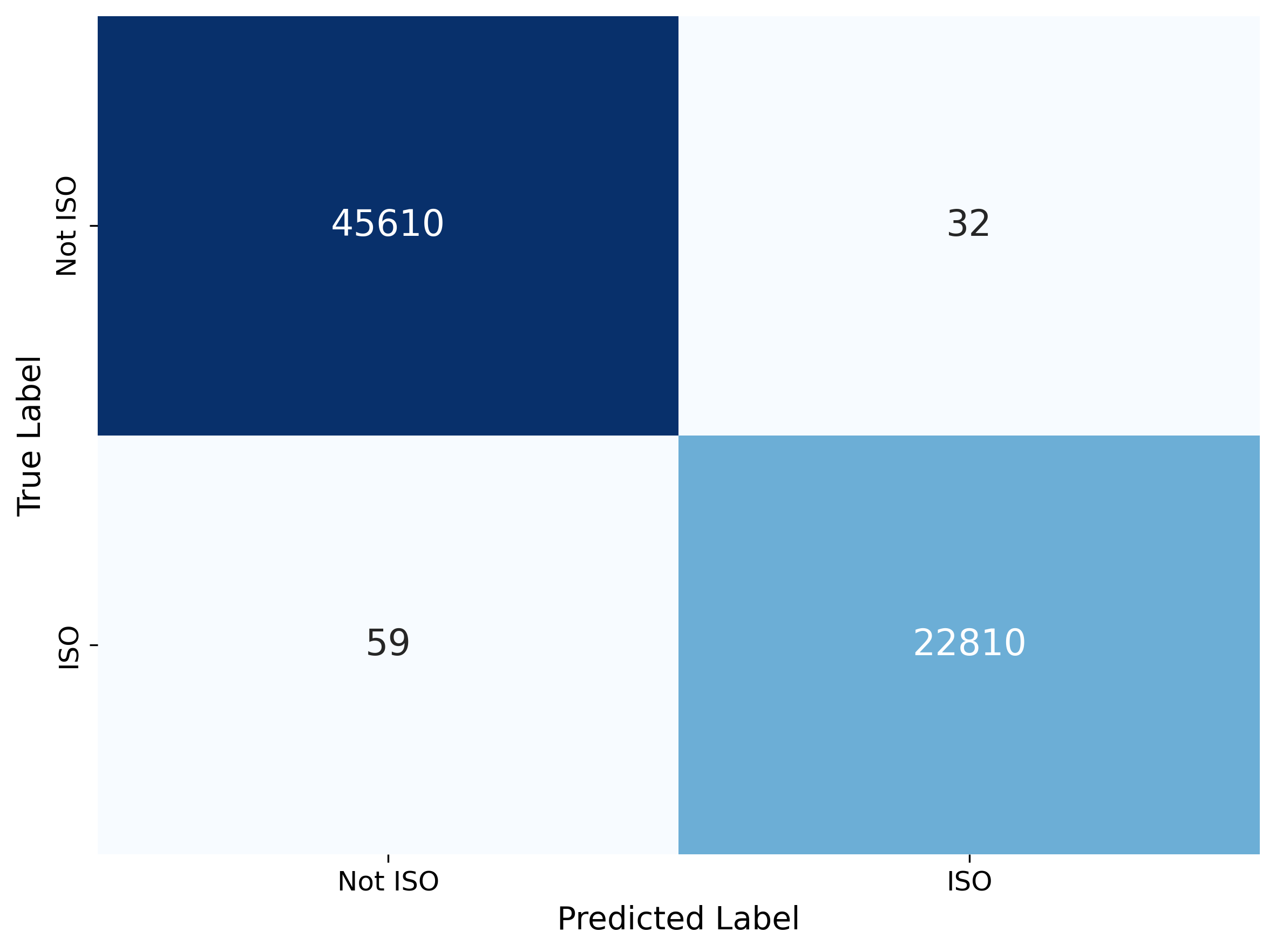}
    \hfill
    \includegraphics[width=0.45\linewidth]{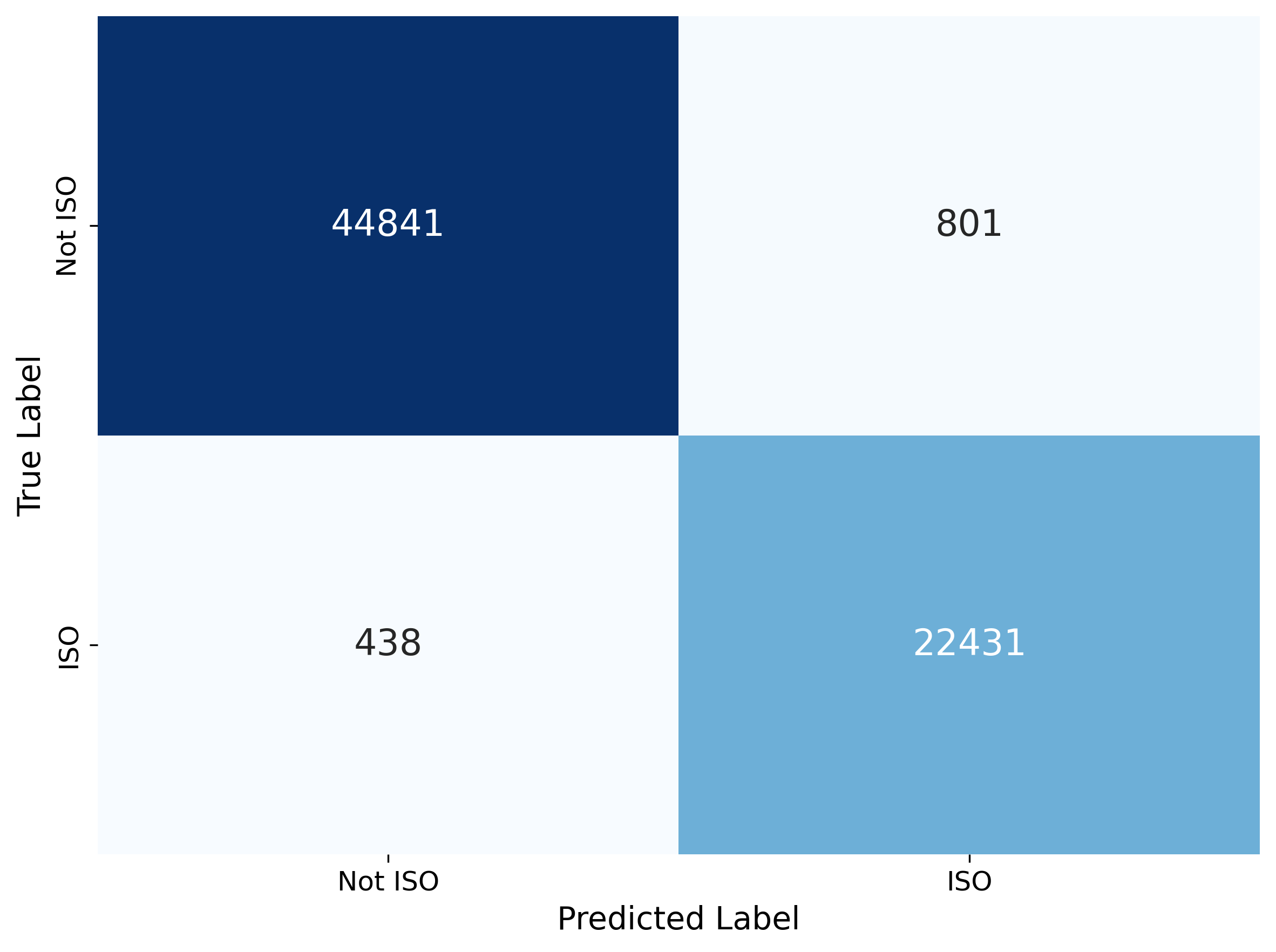}
    \vskip\baselineskip
    \includegraphics[width=0.45\linewidth]{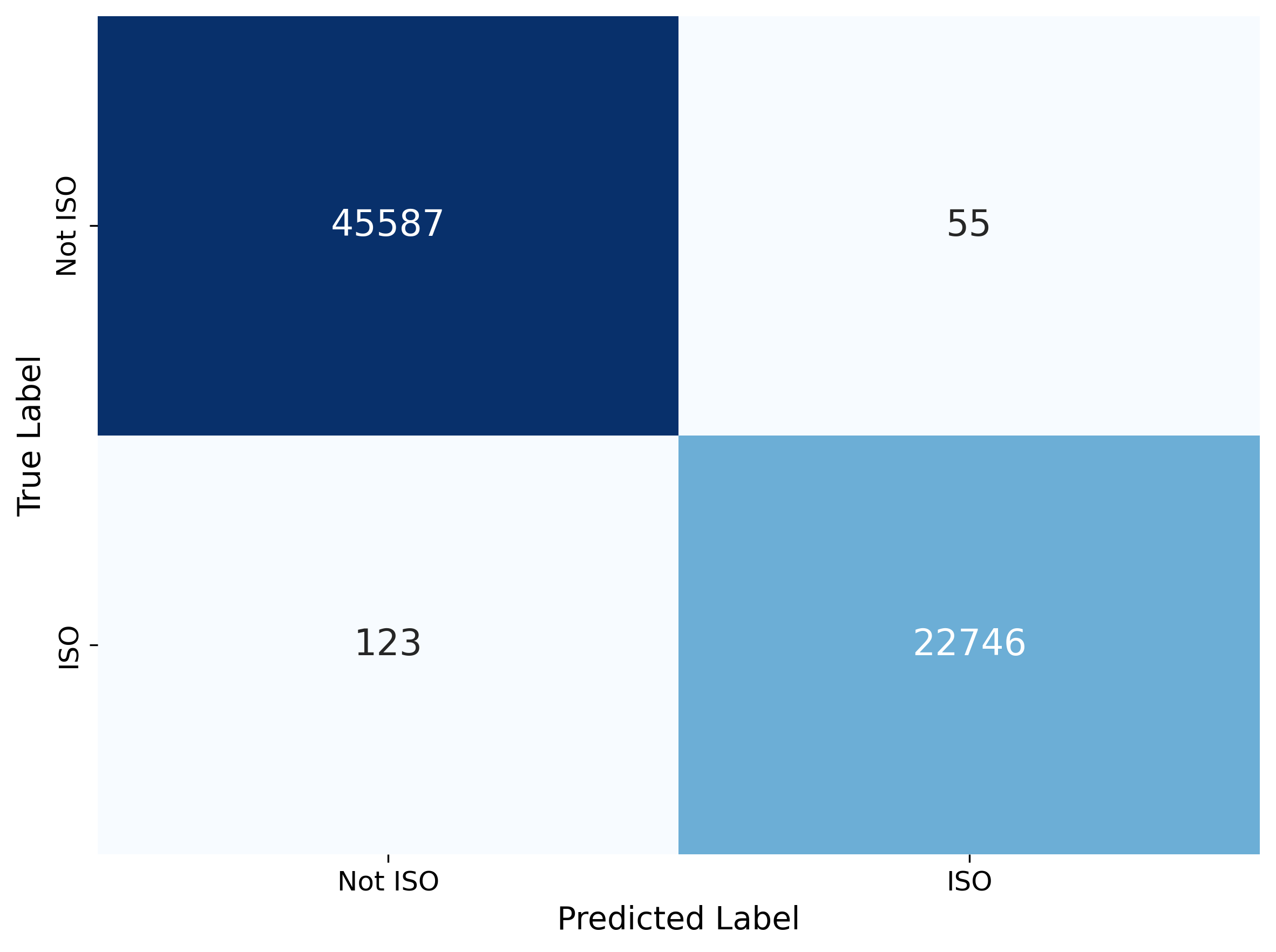}
    \hfill
    \includegraphics[width=0.45\linewidth]{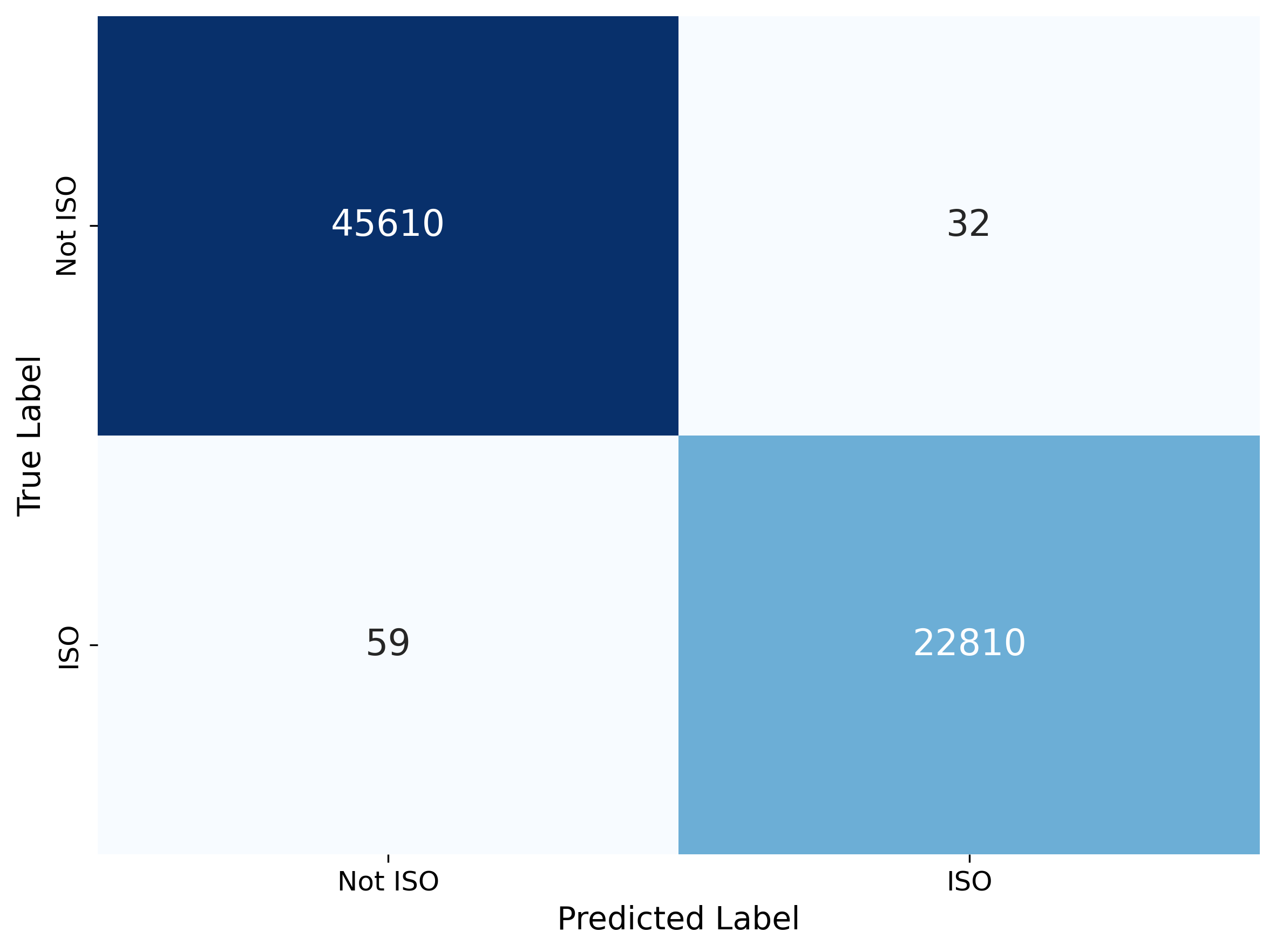}
    \caption{Confusion matrices for GBM (top left), SGD (top right), RF (bottom left), and NN (bottom right).}
    \label{fig:confusion_matrices_grid}
\end{figure*}

Figure~\ref{fig:confusion_matrices_grid} presents the confusion matrices \hl{for our chosen models,} illustrating their classification performance.

The GBM model exhibited high accuracy with minimal false-positives and false-negatives, indicating strong performance in ISO detection. Similarly, the RF model performed well but showed slightly more false-positives. The SGD model  had a higher false-positive rate compared to GBM and RF, suggesting a lower effectiveness in ISO detection. The NN model performed adequately but had a slightly lower accuracy than GBM and RF.

To assess the models' effectiveness in identifying ISOs while minimizing false positives and false negatives, we measured key performance metrics for each model, including F1 score, precision, recall, and accuracy for both ISO and Not-ISO classes. Precision measures the proportion of true ISOs among all objects classified as ISOs by the model. Indeed, a high precision indicates that when the model classifies an object as an ISO, it is likely to be correct. Recall, however, measures the proportion of true ISOs that are correctly identified by the model out of all the actual ISOs in the dataset. A high recall suggests that the model is able to detect a large percentage of the ISOs present.
The F1 score is the harmonic mean of precision and recall, providing a balanced measure of the model's performance. It is particularly useful when the dataset has an uneven class distribution, as is the case with ISOs being rare compared to other celestial objects (although, we ensured that our data were balanced before model training). Accuracy measures the overall correctness of the model's predictions, considering both true positives and true negatives.

As Table~\ref{tab:performance_metrics} illustrates, the GBM and RF models yielded the highest accuracy and balanced performance across all metrics, while the SGD and NN models showed lower accuracy. The results reveals that the GBM model was most effective in distinguishing ISOs from other Solar System objects.

\begin{table*}[ht!]
 \caption{Performance metrics for different models evaluated on the dataset.}
  \centering
  \begin{tabular}{c|c|c|c|c|c}
    \hline
    \textbf{Model}       & \textbf{Class} & \textbf{F1 Score} & \textbf{Precision} & \textbf{Recall} & \textbf{Accuracy}       \\
    \hline
    \multirow{2}{*}{SGD} & NOT ISO        & 0.9864            & 0.9903             & 0.9825                      & \multirow{2}{*}{0.9819} \\
                         & ISO            & 0.9731            & 0.9655             & 0.9808                      &                         \\
    \hline
    \multirow{2}{*}{RF}  & NOT ISO        & 0.9981            & 0.9973             & 0.9988                      & \multirow{2}{*}{0.9974} \\
                         & ISO            & 0.9961            & 0.9976             & 0.9946                      &                         \\
    \hline
    \multirow{2}{*}{GBM} & NOT ISO        & 0.9990            & 0.9987             & 0.9993                      & \multirow{2}{*}{0.9987} \\
                         & ISO            & 0.9980            & 0.9986             & 0.9974                      &                         \\
    \hline
    \multirow{2}{*}{NN}  & NOT ISO        & 0.9884            & 0.9910             & 0.9858                      & \multirow{2}{*}{0.9845} \\
                         & ISO            & 0.9770            & 0.9719             & 0.9821                      &                         \\
    \hline
  \end{tabular}

  \label{tab:performance_metrics}
\end{table*}

\subsection{Validation on nightly datasets}

To assess the performance of our models in a realistic scenario, we validated them on three randomly selected nights from the simulated data, representing typical tracklet counts per year (around 20,000 tracklets per year). Our goal was to minimize confusion and false positives, given the vast quantity of unknown tracklets that LSST will produce. The nights chosen were: 

\begin{itemize}
    \item Night 60313: No ISO, representing a typical night.
    \item Night 60543: Containing 13 ISOs, with a highly exaggerated number of ISOs.
    \item Night 60358: Containing exactly one ISO, simulating a prospective night where a single ISO could be discovered.
\end{itemize}

\begin{table*}[ht!]
\caption{Confusion matrix results for different models evaluated on datasets. Nights: \#60313 (no ISOs), \#60543 (13 ISOs) and \#60358 (one ISO).
}
\centering
\begin{tabular}{l|cccc|cccc|cccc}
\hline
\multirow{2}{*}{\textbf{Model}} & \multicolumn{4}{c|}{\textbf{Dataset \#60313 (No ISO)}} & \multicolumn{4}{c|}{\textbf{Dataset \#60543 (13 ISOs)}} & \multicolumn{4}{c}{\textbf{Dataset \#60358 (One ISO)}} \\
& \textbf{TN} & \textbf{FP} & \textbf{FN} & \textbf{TP} & \textbf{TN} & \textbf{FP} & \textbf{FN} & \textbf{TP} & \textbf{TN} & \textbf{FP} & \textbf{FN} & \textbf{TP} \\
\hline
GBM & 20390 & 19 & 0 & 0 & 21192 & 1 & 0 & 13 & 19887 & 4 & 0 & 1 \\
RF & 20391 & 18 & 0 & 0 & 21190 & 3 & 1 & 12 & 19878 & 13 & 1 & 0 \\
NN & 20346 & 63 & 0 & 0 & 21188 & 5 & 7 & 6 & 19441 & 450 & 0 & 1 \\
SGD & 20343 & 66 & 0 & 0 & 21190 & 3 & 11 & 2 & 19689 & 202 & 0 & 1 \\
\hline
\end{tabular}
\label{tab:confusion_matrix_results_merged}

\end{table*}

The selected data were excluded from the training, testing, and validation datasets, and the models were retrained using the same hyper parameters used in initial training. The results in Table~\ref{tab:confusion_matrix_results_merged} indicate that GBM consistently performed well across various datasets, while other models exhibited varying degrees of effectiveness. A closer look at the GBM results shows that on night 60313, where there were no ISOs in the dataset, the GBM correctly identified 0 ISOs and produced just 19 false positives. On night 60543, the GBM model labeled all 13 ISOs with only one false positive. On night 60358, where one ISO was present, it was correctly identified by the GBM, along with 4 false positives.

\subsection{1I/`Oumuamua and 2I/Borisov}

\begin{table}[ht!]
\caption{Confusion matrix results for different models evaluated on `Oumuamua and Borisov datasets containing tracklets for one night.}
\centering
\begin{tabular}{l|cccc|cccc}
\hline
\multirow{2}{*}{\textbf{Model}} & \multicolumn{4}{c|}{\textbf{`Oumuamua}} & \multicolumn{4}{c}{\textbf{Borisov}} \\
\cline{2-9}
& \textbf{TN} & \textbf{FP} & \textbf{FN} & \textbf{TP} & \textbf{TN} & \textbf{FP} & \textbf{FN} & \textbf{TP} \\
\hline
GBM & 0 & 0 & 45 & 5 & 0 & 0 & 597 & 22  \\
RF & 0 & 0 & 18 & 32 & 0 & 0 & 609 & 103 \\
NN & 0 & 0 & 37 & 13 & 0 & 0 & 516 & 0 \\
SGD & 0 & 0 & 4 & 46 & 0 & 0 & 619 & 0 \\
\hline
\end{tabular}
\label{tab:confusion_matrix_nightly_datasets}
\end{table}

To evaluate the performance of our models on real-world examples of ISOs, we generated two new datasets containing tracklets from the first known ISOs, 1I/`Oumuamua and 2I/Borisov, for a single night~(Table~\ref{tab:confusion_matrix_nightly_datasets}). Despite the challenging nature of the data, the models demonstrated some ability to correctly identify the ISOs.The SGD model achieved the highest true-positive rate for `Oumuamua, correctly identifying 46 out of 50 instances, whereas the RF model performed best for Borisov, correctly identifying 103 out of 712 instances. However, the lack of true-negatives and the presence of false-negatives highlight the difficulty in confidently identifying ISOs from such limited observations. Additionally, the LSST dataset is very differ from current surveys that typically have limiting magnitudes down to +22.5, while LSST will survey significantly deeper (+24.5), thus providing an order of magnitude more unknown faint objects not seen by current surveys. 

 \section{Discussion}
\label{sec:discussion}
Our models demonstrated strong performance on the simulated LSST data, and their application to real-world examples of 1I/`Oumuamua and 2I/Borisov yielded promising results while highlighting areas for improvement. Notably, the models successfully flagged both `Oumuamua and Borisov as potential ISOs without generating any false positives. This achievement is significant as minimizing false positives is crucial to ensure that valuable telescope time is not wasted on follow-up observations of misclassified objects.

The models' ability to correctly identify `Oumuamua and Borisov as ISOs, even with limited observations, underscores their potential for detecting these rare and significant celestial objects. However, the models struggled to achieve high true-positive rates and low false-negative rates for these specific cases, indicating the need for further enhancement to confidently identify ISOs from limited data.

These results emphasize the importance of collecting more comprehensive data on ISOs to improve the models' performance and generalizability. The limited number of observations and the challenging nature of the data pose significant difficulties for the models in confidently identifying ISOs. With only a few detections available for each ISO, the models have limited information to learn from and make accurate predictions. This is evident in the results for 1I/`Oumuamua and 2I/Borisov, where the models struggle to achieve high true-positive rates and low false-negative rates.

To enhance the models' performance in confidently identifying ISOs, several improvements and additional data sources could be considered:

\begin{itemize}
    \item {Collecting more observations of known ISOs:} Increasing the number of observations for confirmed ISOs like 1I/`Oumuamua and 2I/Borisov would provide the models with more examples to learn from, improving their ability to recognize the unique features of ISOs.
    
    \item {Collaborating with other observatories and surveys:} Sharing data and combining observations from multiple telescopes and surveys could help create a more comprehensive dataset of ISO detections, increasing the diversity and quantity of examples available for training the models.
    
    \item {Incorporating additional features:} Extending the feature set to include more physical and morphological properties of ISOs, such as color, spectral characteristics, and light curves, could provide the models with additional discriminating information to improve their classification performance.

    \item \hl{{Investigating the extendedness of ISOs:} Extendedness can affect the quality of astrometry. For instance, typical comets exhibit activity by having a coma or extended tails that can shift the astrometric center, introducing uncertainties in position measurements. Understanding these effects is crucial, as the extended nature of an object can degrade the accuracy of the astrometric data and, consequently, affect the models’ ability to correctly link and identify ISOs. Importantly, moving detections made by LSST that are extended or differ significantly from the stellar point-spread-functions will be flagged immediately and distributed as LSST alerts\footnote{\url{https://lse-163.lsst.io/LSE-163.pdf}} well before any linking is made.}
    
    \item {Exploring transfer learning techniques:} Leveraging knowledge gained from other asteroid and comet detection tasks could help improve the models' performance on the limited ISO data available. Transfer learning techniques could be employed to adapt pretrained models to the specific task of ISO detection.

    \item \hl{{Evaluating uncertainties in the LSST pipeline:} Assessing uncertainties in the LSST pipeline is critical, especially regarding their impact on ISO linking and identification. LSST data's astrometric uncertainties may significantly challenge the differentiation between interstellar and Solar System objects. Even minor positional errors could result in inaccurate orbit determinations or object misclassifications. Furthermore, it is vital to analyze the correlations among orbit fit accuracy, determination, and linking precision. Errors in these areas can cascade through the ISO identification process, potentially complicating detection efforts. Comprehensive examination of these correlations will lead to refined, more reliable models. However, a complete evaluation of LSST uncertainties must await the publication and availability of actual observational data.}

    \item {Continuously updating the models:} As new ISOs are discovered and more data become available, regularly updating the models with the latest observations would help them stay current and improve their performance over time.
\end{itemize}

By addressing these challenges and incorporating additional data and techniques, the models' ability to confidently identify ISOs could be significantly enhanced, enabling a more reliable detection and characterization of these rare and important celestial objects. \section{Conclusion}
\label{s:conclusion}

In this study we explored the application of ML algorithms for the automated classification of ISO tracklets in simulated data from the upcoming LSST survey. Our analysis with RFs shows that the Digest2 values are far more important for classifying ISOs than direct observables supplied by LSST data. The GBM and RF models outperform the SGD and NN models in accurately distinguishing ISOs from other Solar System objects. When evaluated on the simulated data, the GBM model achieved the highest precision, recall, and F1 score, making it the most effective approach for identifying these rare and elusive objects. The models were then applied to three randomly selected nights of LSST data and were able to identify all synthetic ISOs. The SGD and RF models performed best on `Oumuamua and Borisov, respectively. All models produced a relatively low false-positive rate (from a few to a few dozen). However, given that thousands of tracklets will be generated nightly, this low false-positive rate will be manageable. To further improve the performance and generalizability of our models, we have outlined several potential ways forward (see Sect. \ref{sec:discussion}). 

As LSST begins operations, it will generate an unprecedented wealth of data, presenting both challenges and opportunities for the astronomical community. The ability to quickly and accurately identify ISO candidates amidst the vast quantity of tracklets will be crucial for enabling timely follow-up observations and further characterization of these unique objects.

In conclusion, our work lays the foundation for the development of an automated ISO tracklet classification system that can be applied to new data collected in the upcoming LSST era. By developing and implementing efficient and robust classification systems, we can unlock the full potential of LSST in discovering and characterizing these rare and valuable objects, paving the way for advances in our understanding of the materials and processes that shape planetary systems throughout the cosmos. 
\begin{acknowledgements}
This work was supported by the MPC's NASA cooperation agreement funding. We also acknowledge support of the  Oumuamua-Laukien fellowship awarded to the Galileo Project at Harvard University by the Laukien Science Foundation. We also thank to Mario Juri\'c, Melissa Graham, Jake Andrew Kurlander, Siegfried Eggl for help with the simulated LSST data.
\end{acknowledgements}


\begin{thebibliography}{}
\expandafter\ifx\csname natexlab\endcsname\relax\def\natexlab#1{#1}\fi

\end{thebibliography}


\begin{thebibliography}{}
\expandafter\ifx\csname natexlab\endcsname\relax\def\natexlab#1{#1}\fi

\bibitem[{{Bannister} {et~al.}(2017){Bannister}, {Schwamb}, {Fraser},
  {Marsset}, {Fitzsimmons}, {Benecchi}, {Lacerda}, {Pike}, {Kavelaars},
  {Smith}, {Stewart}, {Wang}, \& {Lehner}}]{Bannister17}
{Bannister}, M.~T., {Schwamb}, M.~E., {Fraser}, W.~C., {et~al.} 2017, \apjl,
  851, L38

\bibitem[{{Bergner} \& {Seligman}(2023)}]{Bergner23}
{Bergner}, J.~B. \& {Seligman}, D.~Z. 2023, \nat, 615, 610

\bibitem[{{Bialy} \& {Loeb}(2018)}]{Bialy18}
{Bialy}, S. \& {Loeb}, A. 2018, \apjl, 868, L1

\bibitem[{Bolin {et~al.}(2020)Bolin, Lisse, Kasliwal, Quimby, Tan, Copperwheat,
  Lin, Morbidelli, Abe, Bendjoya, {et~al.}}]{bolin2020characterization}
Bolin, B.~T., Lisse, C.~M., Kasliwal, M.~M., {et~al.} 2020, The Astronomical
  Journal, 160, 26

\bibitem[{{Bolin} {et~al.}(2018){Bolin}, {Weaver}, {Fernandez}, {Lisse},
  {Huppenkothen}, {Jones}, {Juri{\'c}}, {Moeyens}, {Schambeau}, {Slater},
  {Ivezi{\'c}}, \& {Connolly}}]{Bolin18}
{Bolin}, B.~T., {Weaver}, H.~A., {Fernandez}, Y.~R., {et~al.} 2018, \apjl, 852,
  L2

\bibitem[{{Bottke} {et~al.}(2005){Bottke}, {Durda}, {Nesvorn{\'y}}, {Jedicke},
  {Morbidelli}, {Vokrouhlick{\'y}}, \& {Levison}}]{Bottke05}
{Bottke}, W.~F., {Durda}, D.~D., {Nesvorn{\'y}}, D., {et~al.} 2005, \icarus,
  179, 63

\bibitem[{Bottou(2010)}]{bottou2010large}
Bottou, L. 2010, in Proceedings of COMPSTAT'2010: 19th International Conference
  on Computational StatisticsParis France, August 22-27, 2010 Keynote, Invited
  and Contributed Papers, Springer, 177--186

\bibitem[{Breiman(2001)}]{breiman2001random}
Breiman, L. 2001, Machine learning, 45, 5

\bibitem[{{Charnoz} \& {Morbidelli}(2003)}]{Charnoz03}
{Charnoz}, S. \& {Morbidelli}, A. 2003, \icarus, 166, 141

\bibitem[{Chawla {et~al.}(2002)Chawla, Bowyer, Hall, \&
  Kegelmeyer}]{chawla2002smote}
Chawla, N.~V., Bowyer, K.~W., Hall, L.~O., \& Kegelmeyer, W.~P. 2002, Journal
  of artificial intelligence research, 16, 321

\bibitem[{{Cook} {et~al.}(2016){Cook}, {Ragozzine}, {Granvik}, \&
  {Stephens}}]{Cook16}
{Cook}, N.~V., {Ragozzine}, D., {Granvik}, M., \& {Stephens}, D.~C. 2016, \apj,
  825, 51

\bibitem[{{Curran}(2021)}]{Curran21}
{Curran}, S.~J. 2021, \aap, 649, L17

\bibitem[{{Dailey} {et~al.}(2010){Dailey}, {Bauer}, {Grav}, {Myers}, {Mainzer},
  {Mainzer}, {Cutri}, {McMillan}, {Jedicke}, {Denneau}, {Walker}, {Wright}, \&
  {WISE Team}}]{Dailey10}
{Dailey}, J., {Bauer}, J., {Grav}, T., {et~al.} 2010, in American Astronomical
  Society Meeting Abstracts, Vol. 216, American Astronomical Society Meeting
  Abstracts \#216, 409.04

\bibitem[{Denneau {et~al.}(2013)Denneau, Jedicke, Grav, Granvik, Kubica,
  Milani, Vere{\v{s}}, Wainscoat, Chang, Pierfederici,
  {et~al.}}]{denneau2013pan}
Denneau, L., Jedicke, R., Grav, T., {et~al.} 2013, Publications of the
  Astronomical Society of the Pacific, 125, 357

\bibitem[{{Denneau} {et~al.}(2013){Denneau}, {Jedicke}, {Grav}, {Granvik},
  {Kubica}, {Milani}, {Vere{\v{s}}}, {Wainscoat}, {Chang}, {Pierfederici},
  {Kaiser}, {Chambers}, {Heasley}, {Magnier}, {Price}, {Myers}, {Kleyna},
  {Hsieh}, {Farnocchia}, {Waters}, {Sweeney}, {Green}, {Bolin}, {Burgett},
  {Morgan}, {Tonry}, {Hodapp}, {Chastel}, {Chesley}, {Fitzsimmons}, {Holman},
  {Spahr}, {Tholen}, {Williams}, {Abe}, {Armstrong}, {Bressi}, {Holmes},
  {Lister}, {McMillan}, {Micheli}, {Ryan}, {Ryan}, \& {Scotti}}]{Denneau13}
{Denneau}, L., {Jedicke}, R., {Grav}, T., {et~al.} 2013, \pasp, 125, 357

\bibitem[{{Do} {et~al.}(2018){Do}, {Tucker}, \& {Tonry}}]{Do18}
{Do}, A., {Tucker}, M.~A., \& {Tonry}, J. 2018, \apjl, 855, L10

\bibitem[{Engelhardt {et~al.}(2017)Engelhardt, Jedicke, Vere{\v{s}},
  Fitzsimmons, Denneau, Beshore, \& Meinke}]{engelhardt2017observational}
Engelhardt, T., Jedicke, R., Vere{\v{s}}, P., {et~al.} 2017, The Astronomical
  Journal, 153, 133

\bibitem[{Fitzsimmons {et~al.}(2018)Fitzsimmons, Snodgrass, Rozitis, Yang,
  Hyland, Seccull, Bannister, Fraser, Jedicke, \&
  Lacerda}]{fitzsimmons2018spectroscopy}
Fitzsimmons, A., Snodgrass, C., Rozitis, B., {et~al.} 2018, Nature Astronomy,
  2, 133

\bibitem[{{Flekk{\o}y} \& {Brodin}(2022)}]{Flekkoy22}
{Flekk{\o}y}, E.~G. \& {Brodin}, J.~F. 2022, \apjl, 925, L11

\bibitem[{{Flekk{\o}y} {et~al.}(2019){Flekk{\o}y}, {Luu}, \&
  {Toussaint}}]{Flekk19}
{Flekk{\o}y}, E.~G., {Luu}, J., \& {Toussaint}, R. 2019, \apjl, 885, L41

\bibitem[{{Francis}(2005)}]{Francis05}
{Francis}, P.~J. 2005, \apj, 635, 1348

\bibitem[{{Fraser} {et~al.}(2018){Fraser}, {Pravec}, {Fitzsimmons}, {Lacerda},
  {Bannister}, {Snodgrass}, \& {Smoli{\'c}}}]{Fraser18}
{Fraser}, W.~C., {Pravec}, P., {Fitzsimmons}, A., {et~al.} 2018, Nature
  Astronomy, 2, 383

\bibitem[{Friedman(2001)}]{friedman2001greedy}
Friedman, J.~H. 2001, Annals of statistics, 1189

\bibitem[{{Grav} {et~al.}(2011){Grav}, {Jedicke}, {Denneau}, {Chesley},
  {Holman}, \& {Spahr}}]{Grav11}
{Grav}, T., {Jedicke}, R., {Denneau}, L., {et~al.} 2011, \pasp, 123, 423

\bibitem[{He \& Garcia(2009)}]{he2009learning}
He, H. \& Garcia, E.~A. 2009, IEEE Transactions on knowledge and data
  engineering, 21, 1263

\bibitem[{{Heinze} {et~al.}(2022){Heinze}, {Eggl}, {Juric}, {Moeyens}, {Jones},
  {Sullivan}, \& {Bellm}}]{Heinze22}
{Heinze}, A., {Eggl}, S., {Juric}, M., {et~al.} 2022, in AAS/Division for
  Planetary Sciences Meeting Abstracts, Vol.~54, AAS/Division for Planetary
  Sciences Meeting Abstracts, 504.04

\bibitem[{{Hoang} \& {Loeb}(2020)}]{Hoang20}
{Hoang}, T. \& {Loeb}, A. 2020, \apjl, 899, L23

\bibitem[{{Hoang} \& {Loeb}(2023)}]{Hoang23}
{Hoang}, T. \& {Loeb}, A. 2023, \apjl, 951, L34

\bibitem[{{Holman} {et~al.}(2018){Holman}, {Payne}, {Blankley}, {Janssen}, \&
  {Kuindersma}}]{Holman18}
{Holman}, M.~J., {Payne}, M.~J., {Blankley}, P., {Janssen}, R., \&
  {Kuindersma}, S. 2018, \aj, 156, 135

\bibitem[{Hoover {et~al.}(2022)Hoover, Seligman, \& Payne}]{Hoover22}
Hoover, D.~J., Seligman, D.~Z., \& Payne, M.~J. 2022, The Planetary Science
  Journal, 3, 71

\bibitem[{Ivezi{\'c} {et~al.}(2019)Ivezi{\'c}, Kahn, Tyson, Abel, Acosta,
  Allsman, Alonso, AlSayyad, Anderson, Andrew, {et~al.}}]{ivezic2019lsst}
Ivezi{\'c}, {\v{Z}}., Kahn, S.~M., Tyson, J.~A., {et~al.} 2019, The
  Astrophysical Journal, 873, 111

\bibitem[{{Jackson} \& {Desch}(2021)}]{Jackson21}
{Jackson}, A.~P. \& {Desch}, S.~J. 2021, Journal of Geophysical Research
  (Planets), 126, e06706

\bibitem[{Japkowicz \& Stephen(2002)}]{japkowicz2002class}
Japkowicz, N. \& Stephen, S. 2002, Intelligent data analysis, 6, 429

\bibitem[{{Jewitt}(2003)}]{Jewitt03}
{Jewitt}, D. 2003, Earth Moon and Planets, 92, 465

\bibitem[{{Jewitt} {et~al.}(2017){Jewitt}, {Luu}, {Rajagopal}, {Kotulla},
  {Ridgway}, {Liu}, \& {Augusteijn}}]{Jewitt17}
{Jewitt}, D., {Luu}, J., {Rajagopal}, J., {et~al.} 2017, \apjl, 850, L36

\bibitem[{{Jewitt} \& {Seligman}(2023)}]{Je23}
{Jewitt}, D. \& {Seligman}, D.~Z. 2023, \araa, 61, 197

\bibitem[{Jones {et~al.}(2009)Jones, Chesley, Connolly, Harris, Ivezic,
  Knezevic, Kubica, Milani, Trilling, \& Collaboration}]{jones2009solar}
Jones, R., Chesley, S., Connolly, A., {et~al.} 2009, Earth, Moon, and Planets,
  105, 101

\bibitem[{Keys {et~al.}(2019)Keys, Vere{\v{s}}, Payne, Holman, Jedicke,
  Williams, Spahr, Asher, \& Hergenrother}]{keys2019digest2}
Keys, S., Vere{\v{s}}, P., Payne, M.~J., {et~al.} 2019, Publications of the
  Astronomical Society of the Pacific, 131, 1

\bibitem[{{Knight} {et~al.}(2017){Knight}, {Protopapa}, {Kelley}, {Farnham},
  {Bauer}, {Bodewits}, {Feaga}, \& {Sunshine}}]{Knight17}
{Knight}, M.~M., {Protopapa}, S., {Kelley}, M. S.~P., {et~al.} 2017, \apjl,
  851, L31

\bibitem[{LeCun {et~al.}(2015)LeCun, Bengio, \& Hinton}]{lecun2015deep}
LeCun, Y., Bengio, Y., \& Hinton, G. 2015, nature, 521, 436

\bibitem[{{Levine} {et~al.}(2021){Levine}, {Cabot}, {Seligman}, \&
  {Laughlin}}]{Levine21}
{Levine}, W.~G., {Cabot}, S. H.~C., {Seligman}, D., \& {Laughlin}, G. 2021,
  \apj, 922, 39

\bibitem[{{Loeb}(2022)}]{Loeb22}
{Loeb}, A. 2022, Astrobiology, 22, 1392

\bibitem[{{Loeb}(2023)}]{Loeb23}
{Loeb}, A. 2023, Research Notes of the American Astronomical Society, 7, 43

\bibitem[{Louppe {et~al.}(2013)Louppe, Wehenkel, Sutera, \&
  Geurts}]{louppe2013understanding}
Louppe, G., Wehenkel, L., Sutera, A., \& Geurts, P. 2013, Advances in neural
  information processing systems, 26

\bibitem[{Mar{\v{c}}eta \& Seligman(2023)}]{marvceta2023synthetic}
Mar{\v{c}}eta, D. \& Seligman, D.~Z. 2023, The Planetary Science Journal, 4,
  230

\bibitem[{{Masci} {et~al.}(2019){Masci}, {Laher}, {Rusholme}, {Shupe}, {Groom},
  {Surace}, {Jackson}, {Monkewitz}, {Beck}, {Flynn}, {Terek}, {Landry},
  {Hacopians}, {Desai}, {Howell}, {Brooke}, {Imel}, {Wachter}, {Ye}, {Lin},
  {Cenko}, {Cunningham}, {Rebbapragada}, {Bue}, {Miller}, {Mahabal}, {Bellm},
  {Patterson}, {Juri{\'c}}, {Golkhou}, {Ofek}, {Walters}, {Graham}, {Kasliwal},
  {Dekany}, {Kupfer}, {Burdge}, {Cannella}, {Barlow}, {Van Sistine}, {Giomi},
  {Fremling}, {Blagorodnova}, {Levitan}, {Riddle}, {Smith}, {Helou}, {Prince},
  \& {Kulkarni}}]{Masci19}
{Masci}, F.~J., {Laher}, R.~R., {Rusholme}, B., {et~al.} 2019, \pasp, 131,
  018003

\bibitem[{{Mashchenko}(2019)}]{Ma19}
{Mashchenko}, S. 2019, \mnras, 489, 3003

\bibitem[{{Masiero}(2017)}]{Masiero17}
{Masiero}, J. 2017, arXiv e-prints, arXiv:1710.09977

\bibitem[{{McGlynn} \& {Chapman}(1989)}]{McGlynn89}
{McGlynn}, T.~A. \& {Chapman}, R.~D. 1989, \apjl, 346, L105

\bibitem[{Meech {et~al.}(2017)Meech, Weryk, Micheli, Kleyna, Hainaut, Jedicke,
  Wainscoat, Chambers, Keane, Petric, {et~al.}}]{meech2017brief}
Meech, K.~J., Weryk, R., Micheli, M., {et~al.} 2017, Nature, 552, 378

\bibitem[{{Micheli} {et~al.}(2018){Micheli}, {Farnocchia}, {Meech}, {Buie},
  {Hainaut}, {Prialnik}, {Sch{\"o}rghofer}, {Weaver}, {Chodas}, {Kleyna},
  {Weryk}, {Wainscoat}, {Ebeling}, {Keane}, {Chambers}, {Koschny}, \&
  {Petropoulos}}]{Micheli18}
{Micheli}, M., {Farnocchia}, D., {Meech}, K.~J., {et~al.} 2018, \nat, 559, 223

\bibitem[{{Miret-Roig} {et~al.}(2022){Miret-Roig}, {Bouy}, {Raymond}, {Tamura},
  {Bertin}, {Barrado}, {Olivares}, {Galli}, {Cuillandre}, {Sarro}, {Berihuete},
  \& {Hu{\'e}lamo}}]{Miret22}
{Miret-Roig}, N., {Bouy}, H., {Raymond}, S.~N., {et~al.} 2022, Nature
  Astronomy, 6, 89

\bibitem[{{Moro-Mart{\'\i}n} {et~al.}(2009){Moro-Mart{\'\i}n}, {Turner}, \&
  {Loeb}}]{Moro09}
{Moro-Mart{\'\i}n}, A., {Turner}, E.~L., \& {Loeb}, A. 2009, \apj, 704, 733

\bibitem[{{Pe{\~n}a Ram{\'\i}rez} {et~al.}(2012){Pe{\~n}a Ram{\'\i}rez},
  {B{\'e}jar}, {Zapatero Osorio}, {Petr-Gotzens}, \& {Mart{\'\i}n}}]{Pena12}
{Pe{\~n}a Ram{\'\i}rez}, K., {B{\'e}jar}, V.~J.~S., {Zapatero Osorio}, M.~R.,
  {Petr-Gotzens}, M.~G., \& {Mart{\'\i}n}, E.~L. 2012, \apj, 754, 30

\bibitem[{{Portegies Zwart} {et~al.}(2018){Portegies Zwart}, {Torres},
  {Pelupessy}, {B{\'e}dorf}, \& {Cai}}]{Zwart18}
{Portegies Zwart}, S., {Torres}, S., {Pelupessy}, I., {B{\'e}dorf}, J., \&
  {Cai}, M.~X. 2018, \mnras, 479, L17

\bibitem[{{Rafikov}(2018)}]{Rafikov18}
{Rafikov}, R.~R. 2018, \apjl, 867, L17

\bibitem[{{Raymond} {et~al.}(2018){Raymond}, {Armitage}, \&
  {Veras}}]{Raymond18}
{Raymond}, S.~N., {Armitage}, P.~J., \& {Veras}, D. 2018, \apjl, 856, L7

\bibitem[{{Raymond} {et~al.}(2020){Raymond}, {Kaib}, {Armitage}, \&
  {Fortney}}]{Raymond20}
{Raymond}, S.~N., {Kaib}, N.~A., {Armitage}, P.~J., \& {Fortney}, J.~J. 2020,
  \apjl, 904, L4

\bibitem[{{Scholz} {et~al.}(2012){Scholz}, {Muzic}, {Geers}, {Bonavita},
  {Jayawardhana}, \& {Tamura}}]{Scholz12}
{Scholz}, A., {Muzic}, K., {Geers}, V., {et~al.} 2012, \apj, 744, 6

\bibitem[{Schwamb {et~al.}(2023)Schwamb, Jones, Yoachim, Volk, Dorsey, Opitom,
  Greenstreet, Lister, Snodgrass, Bolin, {et~al.}}]{Schwamb23}
Schwamb, M.~E., Jones, R.~L., Yoachim, P., {et~al.} 2023, The Astrophysical
  Journal Supplement Series, 266, 22

\bibitem[{{Sen} \& {Rana}(1993)}]{Sen93}
{Sen}, A.~K. \& {Rana}, N.~C. 1993, \aap, 275, 298

\bibitem[{{Siraj} \& {Loeb}(2022)}]{Si22}
{Siraj}, A. \& {Loeb}, A. 2022, \na, 92, 101730

\bibitem[{Sun {et~al.}(2009)Sun, Wong, \& Kamel}]{sun2009classification}
Sun, Y., Wong, A.~K., \& Kamel, M.~S. 2009, International journal of pattern
  recognition and artificial intelligence, 23, 687

\bibitem[{{Torbett}(1986)}]{Torbett86}
{Torbett}, M.~V. 1986, \aj, 92, 171

\bibitem[{{Trilling} {et~al.}(2018){Trilling}, {Mommert}, {Hora}, {Farnocchia},
  {Chodas}, {Giorgini}, {Smith}, {Carey}, {Lisse}, {Werner}, {McNeill},
  {Chesley}, {Emery}, {Fazio}, {Fernandez}, {Harris}, {Marengo}, {Mueller},
  {Roegge}, {Smith}, {Weaver}, {Meech}, \& {Micheli}}]{Trilling18}
{Trilling}, D.~E., {Mommert}, M., {Hora}, J.~L., {et~al.} 2018, \aj, 156, 261

\bibitem[{{Vere{\v{s}}} \& {Chesley}(2017)}]{Veres17b}
{Vere{\v{s}}}, P. \& {Chesley}, S.~R. 2017, \aj, 154, 13

\bibitem[{{Ye} {et~al.}(2017){Ye}, {Zhang}, {Kelley}, \& {Brown}}]{Ye17}
{Ye}, Q.-Z., {Zhang}, Q., {Kelley}, M. S.~P., \& {Brown}, P.~G. 2017, \apjl,
  851, L5

\end{thebibliography}
\end{document}